\documentclass{aa}
\usepackage{graphicx}
\usepackage{txfonts}
\begin{document}
\title{Depth estimation of the Large and Small Magellanic Clouds}
\author{ Smitha Subramanian, Annapurni Subramaniam}
\institute{Indian Institute of Astrophysics, Bangalore,India\\
            \email{smitha@iiap.res.in, purni@iiap.res.in}}
\date{Received, accepted}
\abstract
 {A systematic estimation of the line of sight depth in the disk and bar
regions of the Large and Small Magellanic Clouds (LMC and SMC)  
using red clump stars is presented.} {We used the red clump stars from the photometric data 
of the Optical Gravitational Lensing Experiment  (OGLE II) survey and 
the Magellanic Cloud Photometric Survey (MCPS) for both the Clouds to estimate the depth. }
{The observed dispersion in the magnitude and colour distribution of red clump stars is used to 
estimate the depth, after correcting for population effects, internal 
reddening within the Clouds and photometric errors.}{The observed 
dispersion due to the line of sight depth ranges from 0.023 mag,
to 0.45 mag (a depth of 500 pc to 10.4 kpc) for the LMC and, from 0.025 to 0.34 magnitude (a depth of 670 pc to 9.53 kpc)
for the SMC. The minimum value corresponds to the dispersion that can be estimated due to errors.
The depth profile of the LMC bar indicates that it is flared. The average depth in the bar region
is 4.0$\pm$1.4 kpc. The northern disk is found to have 
depth (4.17$\pm$0.97 kpc) larger than the southern part of the disk (2.63$\pm$0.8 kpc). 
There is no indication of 
depth variation between the eastern (2.8$\pm$0.92 kpc) and the western (3.09$\pm$0.99 kpc) disk.
The average depth for the disk is 3.44$\pm$ 1.16 kpc.
The SMC is found to have larger depth than the LMC.
In the case of the SMC, the bar depth (4.90$\pm$1.23 kpc) and the disk depth (4.23$\pm$1.48 kpc)
are found to be within the standard deviations.
A prominent feature in the SMC is the increase in depth near the 
optical center. The averaged depth profile near the center resembles a structure like a bulge. }
{The large dispersions estimated in the LMC bar and the northern disk suggest that 
the LMC either has large depth and/or different stellar population in these regions.
The halo of the LMC (using RR Lyrae stars) is found to have larger depth 
compared to the disk/bar, which supports the existence of an inner halo for the LMC.
On the other hand, the estimated depths for the halo (RR Lyrae stars) and disk 
are found to be similar, for the SMC bar region. Thus, increased depth 
and enhanced stellar as well as HI density near the 
optical center suggests that the SMC may have a bulge.}


\authorrunning{Subramanian \& Subramaniam}
\titlerunning{Depth of Magellanic Clouds}
\maketitle

\section{Introduction}
\hspace{0.5cm}
The Large Magellanic Cloud (LMC) and Small Magellanic Cloud (SMC) are 
nearby irregular galaxies at a 
distance of about 50 kpc and 60 kpc respectively. Both these galaxies are nearly face on. 
Magellanic Clouds (MCs) were believed to have had interactions with our Galaxy 
as well as between each other
(Westerlund \cite{westerlund}). It is also believed that the tidal forces due to these interactions have 
caused structural changes in these galaxies. The recent proper motion estimates by Kallivayalil et al
(\cite{K06a}, \cite{K06b}) \& Besla et al (\cite{Besla07}) indicate that these Clouds may be approaching 
our Galaxy for the first time.  These results also claim that the MCs might not
 have always been a binary system. Therefore, it is not clear whether the structure of the 
MCs is modified due 
to their mutual interactions, interactions with our Galaxy or something else, like minor merges.\\

The N-body simulations by Weinberg (\cite{Weinberg}) predicted that the LMC's evolution is 
significantly affected by its interactions with the Milky Way, and the tidal forces will thicken 
and warp the LMC disk. Alves and Nelson (2000) studied the carbon star kinematics and found that 
the scale height, h, increased from 0.3 to 1.6 kpc over the range of radial distance, R, 
0.5 to 5.6 kpc and hence concluded that the LMC disk is flared. Using an expanded sample 
of carbon stars, van der Marel et al. (\cite{van02}) also found that the thickness of the LMC disk increases 
with the radius. The depth of the clouds could vary as a function of radial distance
from the center due to tidal forces. The presence of variation in depth across the LMC, if present,
is likely to give valuable clues to the interactions it has experienced in the past.
 There has not been any direct estimate of the thickness or the line of sight
depth of the bar and disk of the LMC so far.\\

Mathewson, Ford and Visvanathan (\cite{MFV86a} \& \cite{MFV86b}) found that SMC cepheids extend from 
43 to 75 kpc with most cepheids found in the neighbourhood of 59 kpc. Later, the line of sight depth of SMC was 
estimated (Welch \cite{Welch87}) by investigating the line of sight distribution and 
period - luminosity relation of cepheids. They accounted for various factors 
which could contribute to the larger depth estimated by Mathewson et al. (\cite{MFV86a} \& \cite{MFV86b}),
and found the line of sight depth of the SMC to be $\sim$ 3.3 kpc. Hatzidimitriou et al. (\cite{H89}),
estimated the line of sight depth in the outer regions of the SMC to be around 10-20 kpc.\\ 

Red Clump (RC) stars are core helium burning stars, which are metal rich and slightly
more massive counterparts of the
horizontal branch stars. They have tightly defined colour and magnitude, and appear as an
easily identifiable component in colour magnitude diagrams (CMDs). RC stars were used as standard 
candles for distance determination by Stanek et al. (\cite{PS98}). They used the intrinsic 
luminosity to determine the distance to MCs as well as to the bulge 
of our Galaxy.  Olsen and Salyk (\cite{OS02}) 
used their constant I band magnitude to show that the southern LMC disk is warped.
Subramaniam (\cite{S03}) used the constant magnitude of RC stars to show that the LMC has some 
structures and warps in the bar region. 
Their characteristic colour was used by Subramaniam (\cite{S05}) to 
estimate the reddening map towards the central region of the LMC. \\

In this paper, we used the dispersions in the 
colour and magnitude distribution of RC stars for depth estimation. The dispersion in 
colour is due to a combination of observational error, internal reddening (reddening within the disk
of the LMC/SMC) and population effects. The dispersion in magnitude is due to internal 
disk extinction, depth of the distribution, population effects and photometric errors 
associated with the observations. By deconvolving other effects from the dispersion of 
magnitude, we estimated the dispersion only due to the depth of the disk. 
The advantage of choosing RC stars 
as a proxy is that there are large numbers of these stars available to determine the 
dispersions in their distributions with good statistics, throughout the L\&SMC disks.
The depth estimated here would correspond to the depth of the intermediate age L\&SMC disks. 
The depth of the intermediate age disk of these galaxies may give clues to their formation and evolution, and thus
in turn would give clues to their mutual interactions. This could also place some
constraints on their interaction with our Galaxy. 
Measurements of line of sight depth in the central regions of MCs, 
especially the LMC, is of strong interest to understand 
the observed microlensing towards these galaxies.\\

The next section deals with the contribution of population effects to the observed
dispersion of RC stars. Data sources are explained in section 3 and details of the analysis is 
described in section 4. Internal reddening in the MCs is explained in section 5. The LMC and SMC 
results are presented in sections 6 and 7 respectively and their implications are discussed in 
section 8. These results for the disks are compared with the depth estimates of the halo, as 
defined by RR Lyrae stars, in sections 9 and 10. Conclusions are given in section 11.
 
\section{Effect of a heterogeneous population of RC stars} 

The RC stars in the L\&SMC disks are a heterogeneous population and hence, they would
have a range
in mass, age and metallicity. The density of stars in various location will also vary 
with the local star formation rate as a function of time. These factors result in
a range of magnitude and colour of the net population of RC stars in any given location
and would
contribute to the observed dispersion in magnitude and colour distributions. 
Girardi \& Salaris (\cite{GS01}) simulated the RC stars in the LMC using the star formation rate 
estimated by Holtzman et al. (\cite{HGC99}) and the age metallicity relation from 
Pagel and Tautvaisiene (\cite{PT98}). They also simulated the RC stars in SMC using 
star formation results and the age metallicity relation from Pagel and Tautvaisiene (\cite{PT98}). 
The synthetic CMDs of the two systems were obtained and the distribution of RC stars 
is fitted using numerical analysis to obtain the mean and dispersion of the magnitude and
colour distributions. The estimated intrinsic dispersions in 
magnitude and colour distributions for LMC are 0.1 and 0.025 mag respectively. 
In the case of the SMC, the values are 0.076 and 0.03 mag respectively. The values of colour dispersions 
are measured from the CMDs given in the above reference.
We used these estimates of the intrinsic dispersion, to account
for the population effects in our analysis.\\

The above reference used two models of star formation history, one for the bar and the other
for the disk of the LMC. The width of the RC distribution for the above two populations is not very 
different. They also simulated the RC distribution found by Dolphin (\cite{D00}) in the northern LMC. 
The RC stars in this region were found to be very different due to a significantly different star formation
history and metallicity. The effect on the width of the RC was found to be large. The intrinsic dispersion in this region 
was found to be 0.218 mag, double that of the bar population. Therefore, if the population of a region 
is different, its effect on the width of the RC distribution will be to increase it. All the regions 
in the bar and the disk were corrected for population effects using the
bar and the disk model estimates of width. There may be an uncorrected component due to
population effects in the estimated depth, due to variations in the RC population between regions. 
The final estimated variation of RC width will be a combination of variation in depth and population
across the LMC. In the SMC also, there can be an effect of different RC populations in the 
dispersion corresponding to depth. The contribution of these components in various locations of the MCs 
will be discussed later.
       
\begin{figure}
\resizebox{\hsize}{!}{\includegraphics{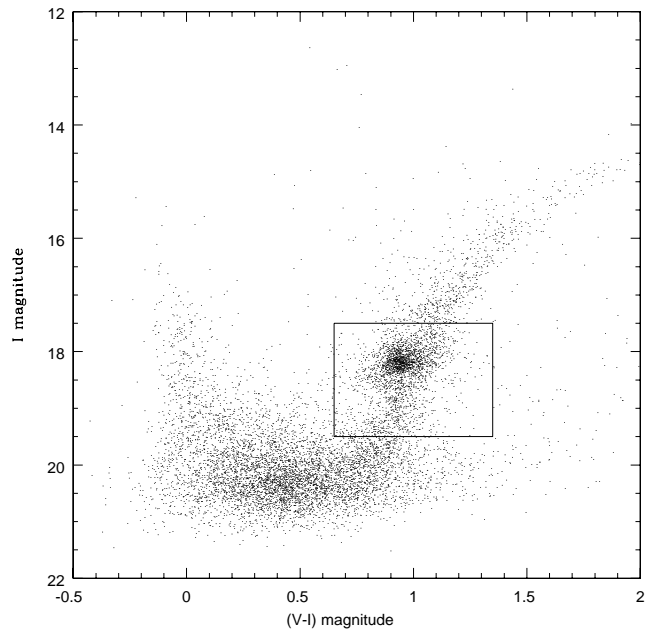}}
\caption{Colour magnitude diagram of an LMC region. The box used to identify the RC
population is shown.}
\end{figure}

\begin{figure}
\resizebox{\hsize}{!}{\includegraphics{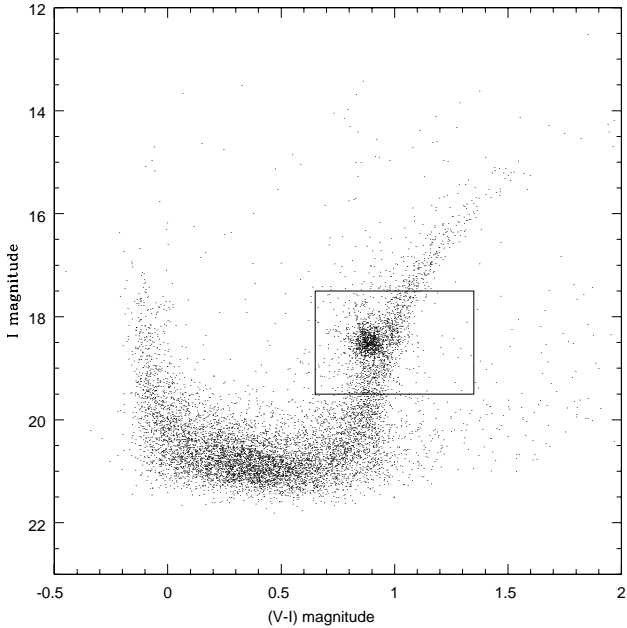}}
\caption{Colour magnitude diagram of an SMC region. The box used to identify the RC
population is shown.}
\end{figure}

\section{Data}
\vspace{0.25cm}
\subsection{LMC data}
The OGLE II survey (Udalski et al. \cite{Uda00}) scanned the central region of the LMC to detect microlensing
events. One of the outcomes of this survey is a  catalogue of stars in the central/bar region of the LMC, 
consisting of photometric data of 7 million stars in the B,V and I pass bands. This catalogue presents data 
of 26 fields which cover the central 5.7 square degrees of the LMC in the sky. Out of the 26 regions/strips, 
21 regions are within 2.5 degrees of the optical center of the LMC representing the bar region, 
and the other 5 regions are in the north western part of the LMC disk. The average photometric error 
of red clump stars in I and V bands are $\sim$ 0.05 magnitude. Photometric data with errors less 
than 0.15 mag are considered for the analysis. 
Each strip was divided into 4$\times$16 smaller regions, each having an area of 3.56$\times$3.56 square arcmin. 
Thus 26 strips of the LMC were divided into 1664 regions. (V$-$I) vs I CMDs were plotted for each region and a sample CMD of one such 
region is shown in figure 1. For all the regions, red clump stars were well within a box in the CMD, with widths
0.65 - 1.35 mag in (V$-$I)colour and 17.5 - 19.5 mag in I magnitude. Thus, the red clump stars 
were identified in each region. The OGLE II data suffers from incompleteness due to crowding effects and the incompleteness 
in the RC distribution is corrected using the values given in Udalski et al. (\cite{Uda00}).\\

The Magellanic Cloud Photometric Survey (MCPS, Zaritsky et al. \cite{ZHT04}) of the central 64 square
degrees of the LMC contains photometric data of around 24 million stars in the U,B,V and I pass
bands. Data with errors less than 0.15 mag are taken for the analysis. The regions away from
the bar are less dense compared to the bar region. The total observed regions are divided into
1512 sub-regions each having an area of approximately 10.53$\times$15 square arcmin. Out of 1512 
regions only 1374 regions have a reasonable number of RC stars to do the analysis. (V-I) vs I CMDs 
for each region were plotted  and red clump stars were identified as described above.\\

\subsection{SMC data}
The OGLE II survey (Udalski et al \cite{Uda98}) of the central region of the SMC contains photometric data 
of 2 million stars in the B,V and I pass bands. The catalogue of SMC presents data of 
11 fields which cover the central 2.5 square degrees of the SMC in the sky. 
The observed regions 
of the SMC are divided into 176 regions. Each strip was divided into 2$\times$8 regions, 
each having an area of 7.12$\times$7.12 square arcmin to obtain enough stars 
in each region. 
Data selection and analysis is similar to that for the LMC, including the box
used to identify the RC stars in the CMD.
A sample CMD for one location is shown in Figure 2. 
We used the incompleteness corrections given in Udalski et al. (\cite{Uda98}).\\

The Magellanic Cloud Photometric Survey (MCPS, Zaritsky et al. \cite{ZHT02}) of the central 18 square 
degrees of the SMC contains photometric data of around 5 million stars in the U,B,V and I pass 
bands. Data with errors less than 0.15 mag are taken for the analysis. The regions away from 
the bar are less dense compared to the bar region. The total observed regions are divided into 
876 sub-regions each having an area of approximately 8.9$\times$ 10 square arcmin. Out of 876  
regions, 755 regions with a reasonable number of RC stars 
were considered for analysis.\\

\begin{figure}
\resizebox{\hsize}{!}{\includegraphics{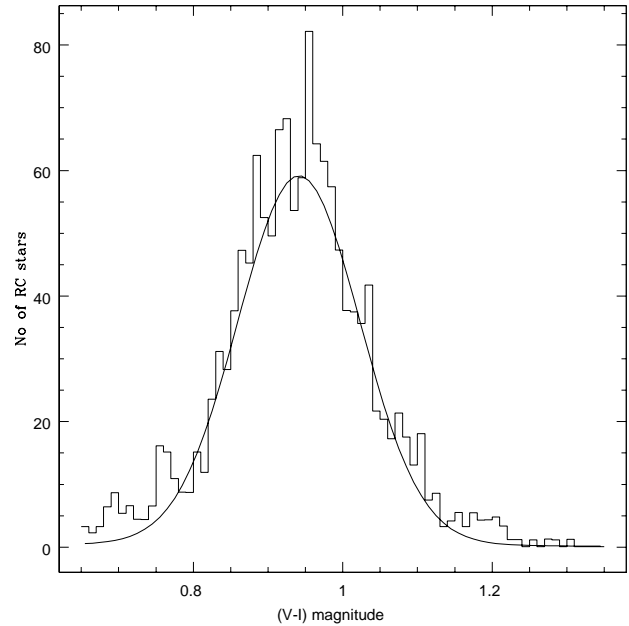}}
\caption{A typical colour distribution of red clump stars in the LMC. The best fit to the distribution is also shown. 
The reduced $\chi^2$ value of this fit is 1.33.}
\end{figure}

\begin{figure}
\resizebox{\hsize}{!}{\includegraphics{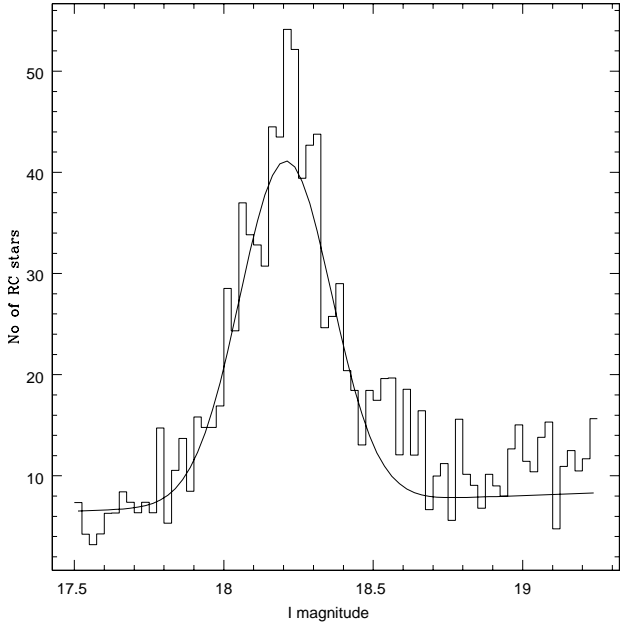}}
\caption{A typical magnitude distribution of red clump stars in the LMC. The best fit to the distribution is also shown.
The reduced $\chi^2$ value of this fit is 0.99.}
\end{figure}

\section{Analysis}
\hspace{0.5cm}
A spread in magnitude and colour of red clump stars is observed in the CMDs of both 
the LMC and SMC. Their number distribution profiles roughly 
resemble a Gaussian. The width of the Gaussian in the distribution of colour is 
due to the internal reddening, apart from observational error and population effects.
The width in the distribution of magnitude is due to population effects, observational error, internal 
extinction and depth. By deconvolving the effects of observational error, extinction and population
effects from the distribution of magnitude, an estimate of depth can be obtained.\\

       To obtain the number distribution of the red clump stars in each region,
the data are binned with a bin size of 0.01 and 0.025 mag in 
colour and magnitude respectively. The obtained distributions in colour and magnitude are 
fitted with a function, a Gaussian + quadratic polynomial. The Gaussian represents the red 
clump stars and the other terms represent the red giants in the region. A non linear least 
square method is used for fitting and the parameters are obtained. 
In figures 3 and 4, the distribution as well as the fitted curve are shown for both colour and 
magnitude distribution of an LMC region (OGLE II data). Similarly for SMC (OGLE II data), the
distribution as well as the fitted curve are shown for both colour and magnitude in figures 5 and 6
. The parameters obtained 
are the coefficients of each term in the function used to fit the profile, the error in the 
estimation of each parameter and the goodness of the fit, which is the same as the reduced $\chi^2$ value. 
Regions with reduced $\chi^2$ values greater than 2.6 are omitted from the analysis. As the 
important parameter for our calculations is the width associated with the two distributions, 
we also omitted regions with fit errors of width greater than 0.1 mag from our analysis.
After these omissions,
the number of regions useful for analysis in LMC (OGLE II data) and LMC (MCPS data) is reduced from 1664 to 1528
and from 1374 to 1301 respectively. Similarly for the SMC, after omitting regions with larger 
reduced $\chi^2$ values and fit error values, the number of regions useful for analysis 
in OGLE II data and MCPS data is reduced
from 176 to 150 and from 755 to 600 respectively. 
Thus, the total observed dispersion in
(V$-$I) colour and I magnitude were estimated for RC stars in all these regions.
The number of RC stars in each region studied in the MCs depends 
on the RC density. The number is large in the central regions, whereas it decreases
in the disk. In LMC, the RC stars range between 500 - 2000 in the bar region, whereas
the range is 200 - 1500 in the disk. In the central regions of the SMC,
the RC stars range between 1000 - 3000. The disk is found to have a range 200 - 1500.\\ 

\begin{figure}
\resizebox{\hsize}{!}{\includegraphics{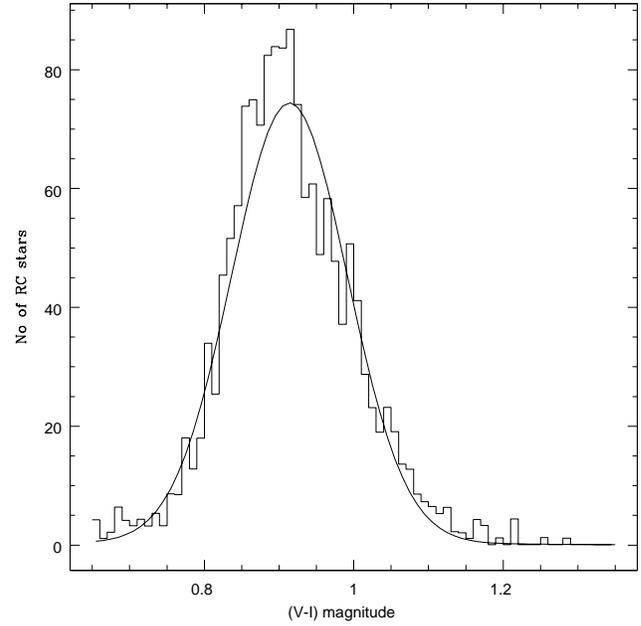}}
\caption{A typical colour distribution of red clump stars in the SMC. The best fit to the distribution is also shown.
The reduced $\chi^2$ value of this fit is 1.18.}
\end{figure}

\begin{figure}
\resizebox{\hsize}{!}{\includegraphics{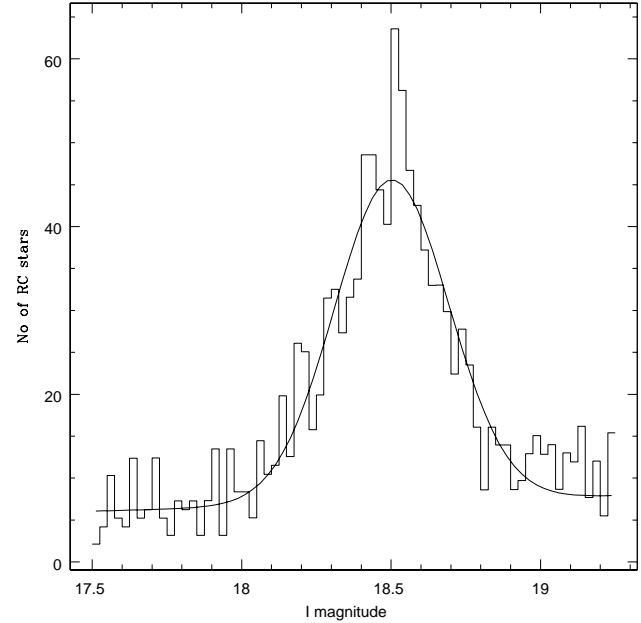}}
\caption{A typical magnitude distribution of red clump stars in the SMC. The best fit to the distribution is also shown.
The reduced $\chi^2$ value of this fit is 1.36.}
\end{figure}    

The following relations are used to estimate the resultant dispersion due to depth from the above
estimated dispersions.

$\sigma^2_{col}$ = 
$\sigma^2_{internal reddening}$ + 
$\sigma$$^2$$_{intrinsic}$ + $\sigma$$^2$$_{error}$\\ \\
$\sigma$$^2$$_{mag}$ = $\sigma$$^2$$_{depth}$ + 
$\sigma$$^2$$_{internal extinction}$ +
$\sigma$$^2$$_{intrinsic}$ + $\sigma$$^2$$_{error}$\\

The average photometric errors in I and V band magnitudes were calculated for each region 
and the error in I magnitude and (V$-$I) colour were estimated. These were subtracted from the observed
width of magnitude and colour distribution respectively, thus accounting for the photometric errors 
(last term in the above equations). The contribution from the heterogeneous population of RC
stars were discussed in section 2, and the dispersion in colour and magnitude due to this
effect ($\sigma_{intrinsic}$) were also subtracted from the observed dispersions.
After correcting for the population effects and the observational error in colour, the remaining
spread in colour distribution (first equation) is taken as due to the internal reddening, E(V$-$I). 
This is converted into extinction in I band using the relation A(I) = 0.934 E(V$-$I), where E(V$-$I) is the 
internal reddening  estimated for each location. This was used to deconvolve the effect of 
internal extinction from the spread in magnitude. The above relation is derived from the relations 
E(V$-$I) = 1.6 E(B$-$V) and  A(I) = 0.482 A(V) (Rieke and Lebofsky \cite{RF85}). The interstellar extinction law of our Galaxy is 
adopted for the calculations of Magellanic Clouds based on the results of the studies by Nandy 
and Morgan (\cite{NM78}), Lequeux et al (\cite{L1982}) and Misslet, Clayton and Gordon (\cite{MCG99}), which showed that both 
LMC and SMC have extinction curves qualitatively similar to those found in Milky Way.\\ 

Thus, the net dispersion  in magnitude due to depth alone was estimated for considered
regions in the LMC and SMC. The resultant width in magnitude is converted into depth in kpc using the
distance modulus formula and taking a distance of 50 kpc to the LMC and a distance of 60 kpc to the SMC.\\

The error in the estimation of the dispersion corresponding to depth is obtained from the errors 
involved in the estimation of width of colour and magnitude distribution. The random 
error associated with the width corresponding to depth is $\Delta$depth$^2$ = 
$\Delta$I$^2$$_w$$_i$$_d$$_t$$_h$ + $\Delta$(V-I)$^2$$_w$$_i$$_d$$_t$$_h$. Thus the associated error 
in the estimation of depth is calculated for all the locations. This error will also translate
as the minimum depth that can be estimated. The minimal depth that can be 
estimated  is $\sim$ 360 pc in the central regions of the LMC and $\sim$
650 pc in the outer regions of the LMC. In the SMC, the minimal thickness that can be estimated 
is $\sim$ 350pc in the central regions of the SMC and $\sim$ 670pc in the 
outer regions of the SMC.\\

\section{Internal reddening in the MCs}
\begin{figure*}
\resizebox{\hsize}{!}{\includegraphics{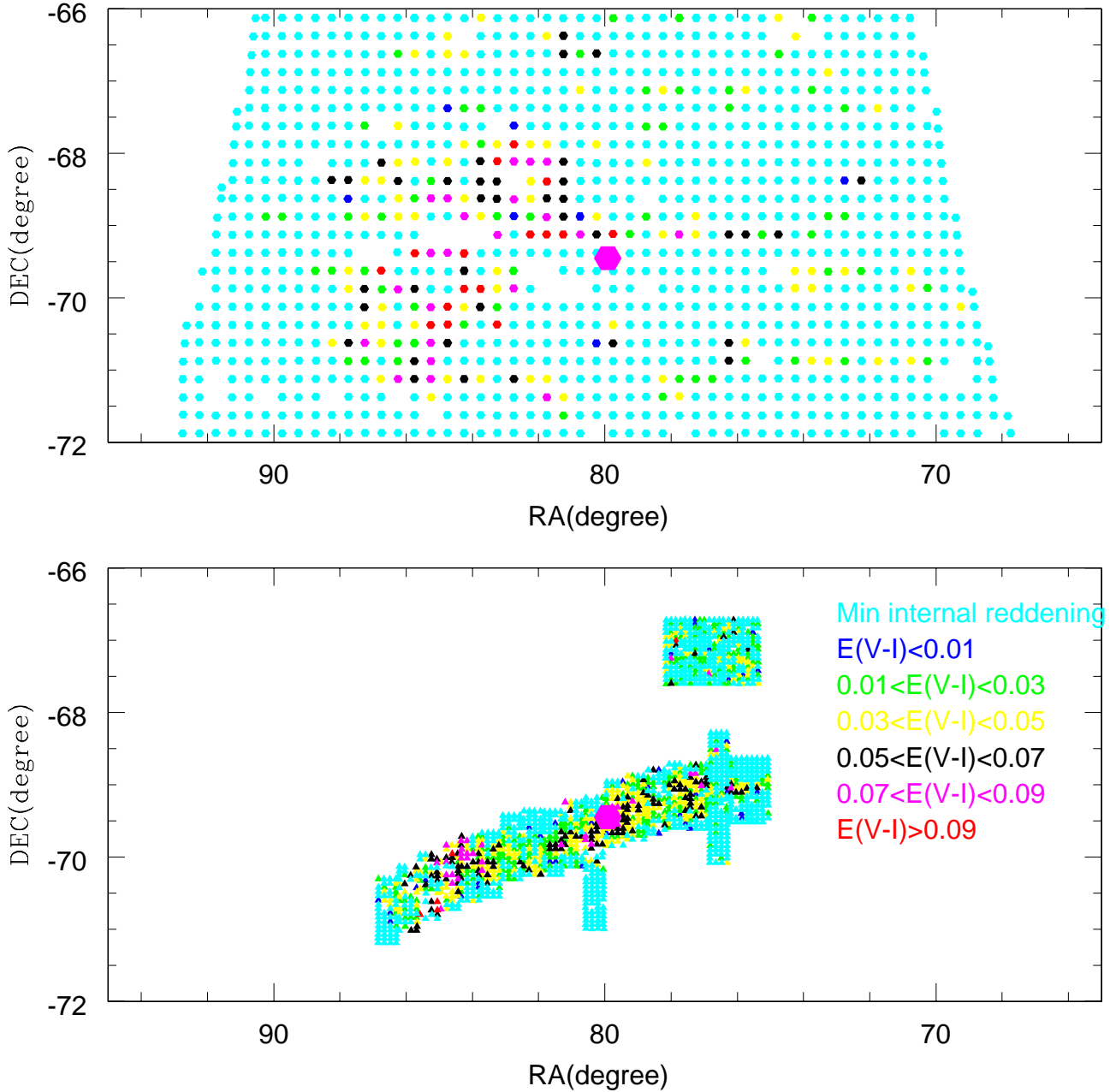}}
\caption{Two dimensional plot of the internal reddening in the LMC. The colour code is given in the figure. The magenta dot 
represents the optical center of the LMC. The upper plot is derived from the MCPS data, whereas the lower plot is
derived from the OGLE II data.}
\end{figure*}
\begin{figure*}
\resizebox{\hsize}{!}{\includegraphics{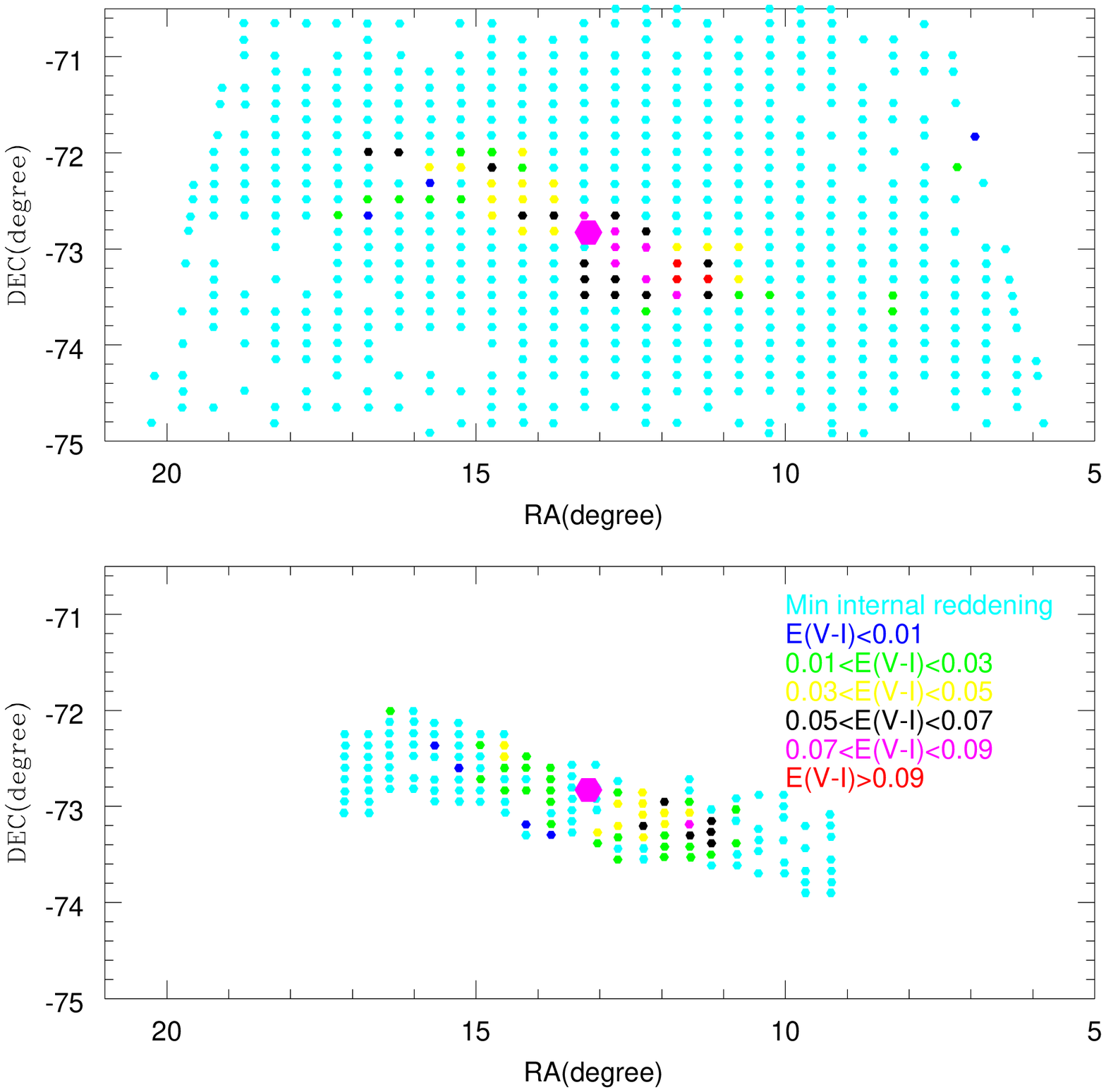}}
\caption{Two dimensional plot of the internal reddening in the SMC. The colour code is given in the figure. The magenta dot 
represents the optical center of the SMC. The upper plot is derived from the MCPS data, whereas the lower plot is derived
from the OGLE II data.}
\end{figure*}
One of the by products of this study is the estimation of internal reddening in the MCs.
The shift of the peak of the (V$-$I) colour distribution from the expected value was used by Subramaniam (\cite{S05}) to
estimate the line of sight reddening map to the OGLE II region of the LMC. The above study
estimated the reddening between the observer and the observed region in the LMC. In this study,
we used the width of the (V$-$I) colour distribution to estimate the internal reddening map across the MCs.
This estimates the front to back reddening of a given region in the MCs, which we call as the internal
reddening (in E(V$-$I)), and does not estimate the reddening between  the front end of the region
and the observer. Thus, this estimate traces the reddening within the bar/disk of the MCs and hence the location of the dust. 
The estimates and figures given below thus gives the internal reddening within the MCs.\\

The colour coded figures of the internal reddening in the LMC and SMC are presented in figure 7 and 8 respectively.
It can be seen that the internal reddening is high only in some specific regions in both the MCs. Most of the
regions have very negligible internal reddening suggesting that most of the regions in the MCs are optically thin.
The regions of high internal reddening in the LMC are located near the eastern end of the bar and the 30 Dor star forming
region. The highest internal reddening estimated is E(V$-$I) = 0.13 mag in the OGLE II region of the LMC,
which is the bar region and  0.16 mag in the MCPS region, close to the 30 Dor location. It is noteworthy
that these are not  very high values. 
The OGLE II internal reddening map shows that, apart from the eastern end of the bar, some regions near the
center also have internal reddening. The MCPS data shows that the 
internal reddening across the LMC disk as seen by the RC stars is very small. The error associated with the 
width of colour distribution is translated as the minimum internal reddening that can be estimated. The minimum internal reddening 
that can be estimated in the central regions as well as in the disk of LMC is 0.003 mag.   
In the case of the SMC, a region of high internal reddening is found to the west of the
optical center. Also, the bar region is found to have some internal reddening, whereas the disk has very little
internal reddening (within the area studied). The highest reddening estimated is E(V$-$I) = 0.08 mag
in the OGLE II regions and 0.12 mag in the MCPS region. These regions are located close to the
optical center. The rest of the bar as well as the disk have very little  internal reddening. Thus,
our results indicate small extinction across the SMC, as seen by the RC stars. The minimum internal reddening that
can be estimated is 0.002 mag in the central regions of SMC and 0.005 mag in the disk of SMC.\\

It is interesting to see low internal reddening across the MCs, as seen by the RC stars. This is in contradiction
to the large extinction expected near the star forming regions, especially near 30 Dor. The reddening is estimated here using the same type of stars that are used to estimate the depth. Thus, we maintain the consistency of the
same tracer for the above two properties. The higher values for the foreground reddening estimates were 
obtained by Harris et al (\cite{Harris97}) using the MCPS data for OB stars, whereas lower values were obtained using RC stars for the same regions
by Subramaniam (\cite{S05}) using OGLE II data. Thus, the reddening values estimated varied with respect to the tracer used.
Once again, the present results confirm that the RC disk has much less internal extinction. This variation
of reddening as a function of the population is suggestive of population segregation across the LMC. The above
results suggest that the star forming regions in the LMC are likely to be behind the RC disk.

\begin{figure*}
\resizebox{\hsize}{!}{\includegraphics{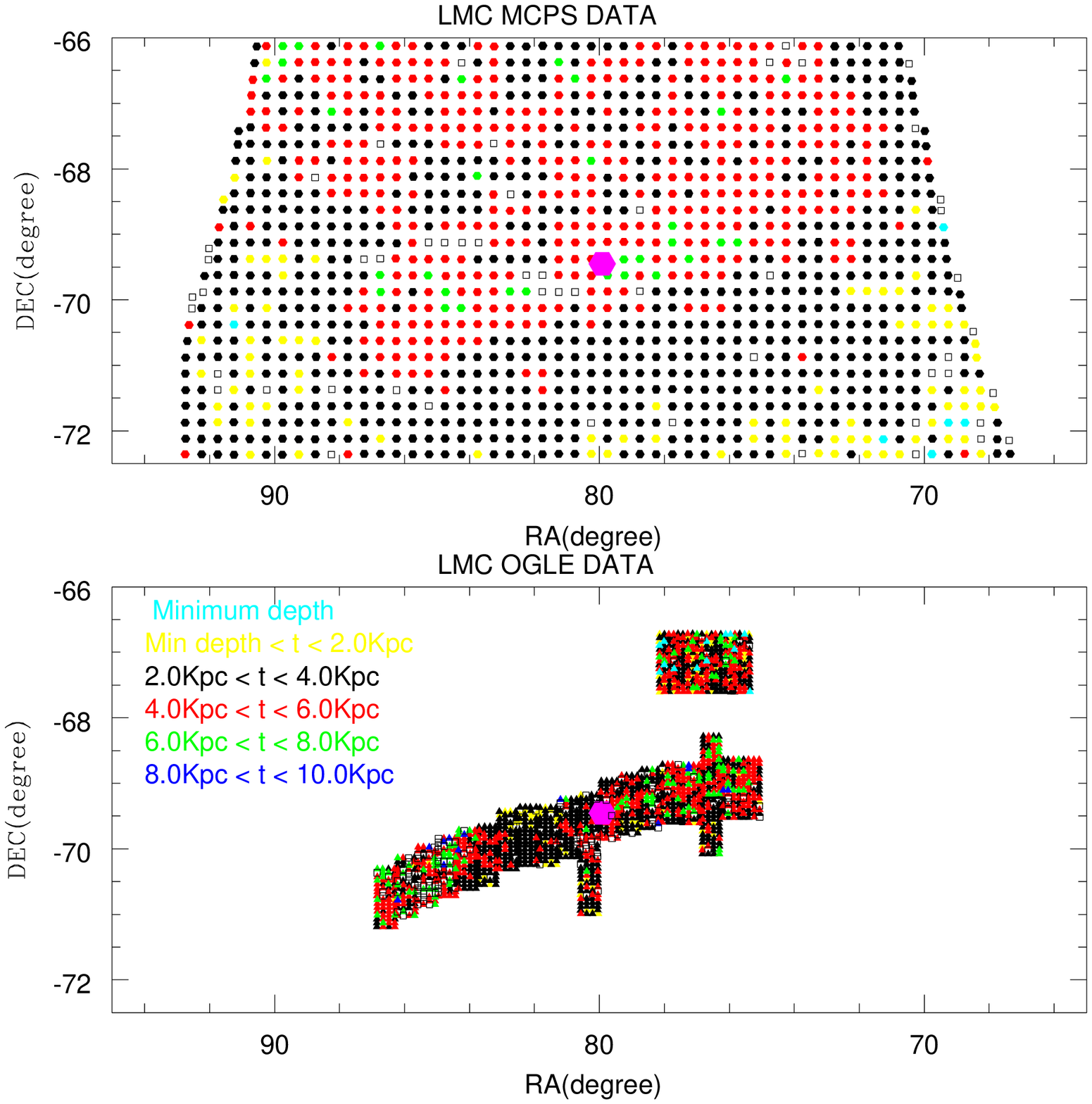}}
\caption{Two dimensional plot of depth (t) in the LMC. The colour code 
is given in the figure. The magenta dot represents the optical center of the LMC.The empty squares represent
the omitted regions with poor fit.}
\end{figure*}

\begin{figure}
\resizebox{\hsize}{!}{\includegraphics{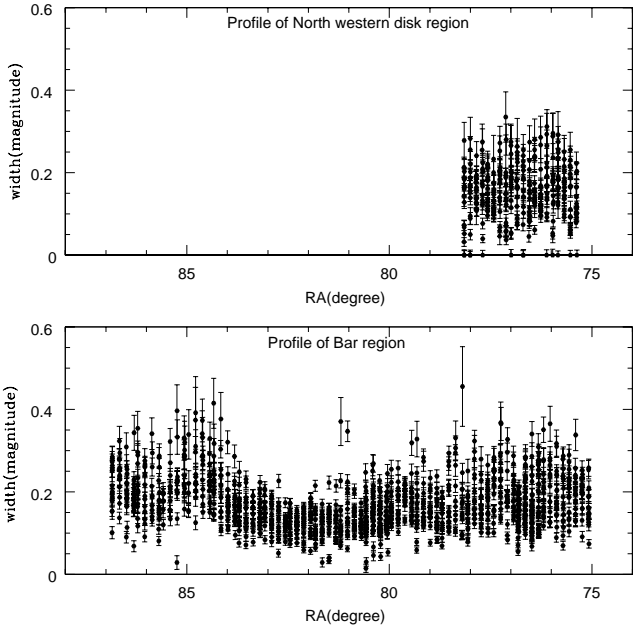}}
\caption{Width corresponding to depth with error bars plotted against RA for both the central 
bar region and north western disk region of LMC.}
\end{figure}

\begin{figure}
\resizebox{\hsize}{!}{\includegraphics{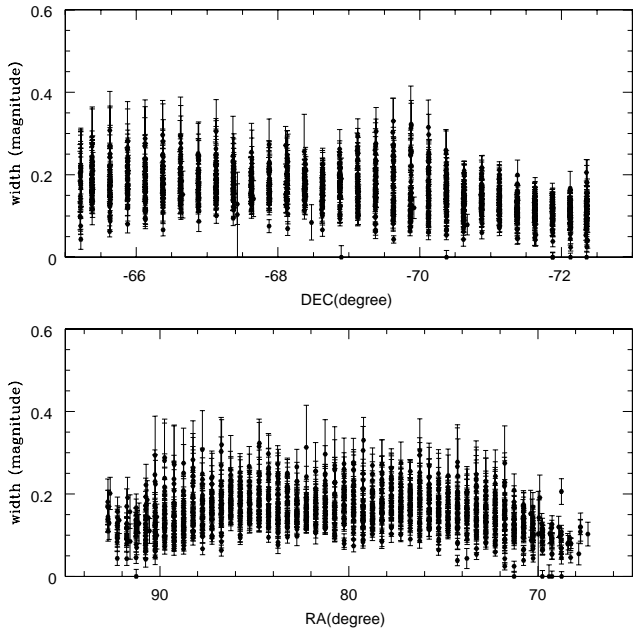}}
\caption{Width corresponding to depth with error bars plotted against RA and Dec for the LMC
MCPS data.}
\end{figure}

\begin{figure}
\resizebox{\hsize}{!}{\includegraphics{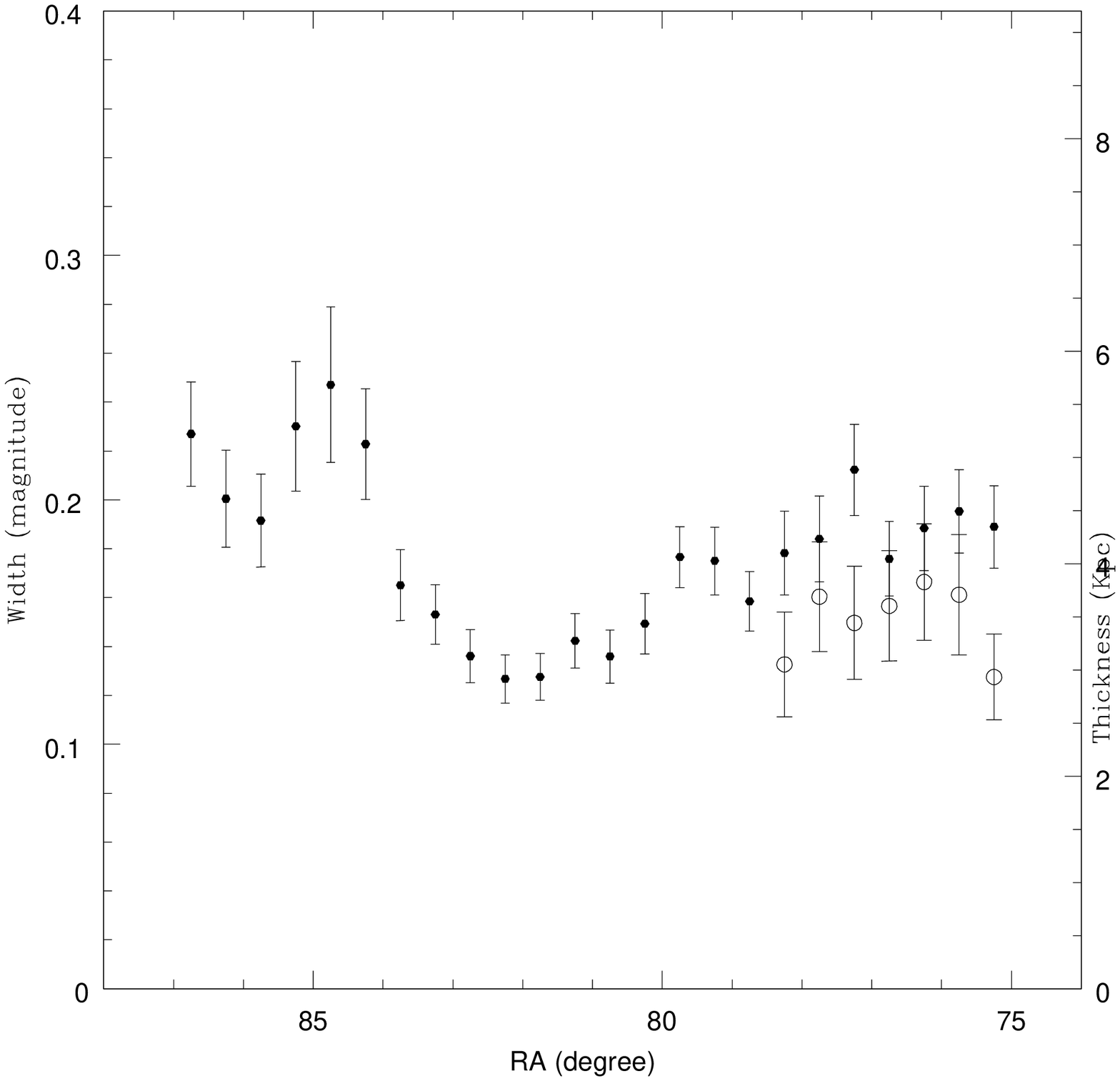}}
\caption{Width corresponding to depth averaged along the declination plotted against RA for 
both the central bar region (closed circles) and north western disk region (open circles)of LMC.}
\end{figure}   

\section{Results : The LMC }
The line of sight depth in the LMC has been derived using two data sets. OGLE II data cover the central
region and this is basically the bar region of the LMC. OGLE II data also cover a small detached disk 
region in the north-western direction. On the other hand, MCPS data cover a
significant area of the disk, besides the bar. Thus the OGLE II data is suitable to derive the 
depth of the bar region, whereas the MCPS data is suitable to derive the depth of the bar as well
as the disk regions. The two data sets can be used for consistency checks and the results
derived will be compared.\\

The depth of 1528 regions of LMC (OGLE II data) were calculated. Out of 1528 regions,
1214 are in the central bar 
region and the remaining 314 regions are in the north western disk region. 
In the north western disk 
region there are regions with minimal thickness. The depth of 1301 regions of LMC (MCPS data) are 
also calculated. Regions with minimal thickness are seen in the disk of LMC.\\

A two dimensional plot of the depth for the 1528 OGLE II regions is shown in the lower panel of figure 9. 
This plot is colour coded as explained. 
The optical center of the LMC is taken to be RA = 05$^h$ 19$^m$ 38$^s$, 
Dec = -69$^0$ 27' 5".2 (J2000.0,de Vaucoulers and Freeman, \cite{dVF73}). 
The OGLE II data show a range of dispersion values from 0.03 to 0.46 mag (a depth of
of 690 pc to 10.5 kpc;  avg: 3.95 $\pm$ 1.42 kpc) for the LMC central bar region. For the N-W disk region,
the dispersion estimated ranges from 0.023 mag to 0.33 mag (a depth of 500 pc to 7.7 kpc;
avg: 3.56 $\pm$ 1.04 kpc).The minimum value in the N-W disk region is limited by errors. The fraction
of such regions, where the minimum value is limited by errors, is 4.06\%.
Regions in the bar between
RA 80 - 84 degrees show a reduced depth (0.5 - 4 kpc, as indicated by yellow and black points). 
The regions to the east and west of the above region are found to have larger depth (2.0 - 8.0 kpc, 
black, red and green points). Thus, the depth of the bar at its ends is greater than that near 
its center. Since the center of the bar does not coincide with the optical center, the thinner
central bar region is located to the east of the optical center. The average values of these three
regions of the bar are 3.21$\pm$ 1.03 kpc for the central minimum, 4.13$\pm$1.35 kpc for the western region
and 4.95$\pm$1.49 kpc for the eastern region. 
This is better brought out in figure 10, where the depth is shown as a function of RA (the Dec range
in OGLE II data is less when compared to the RA range).  The lower panel shows the bar region and the
upper panel shows the N-W disk region. An indication of flaring is seen here. In this figure,
the depth of all the regions are shown and the error bar on each point denotes the error in the estimation
of depth at each location. The depth averaged along the Dec of each RA is shown in figure 12. This plot 
clearly suggests that the bar is flared at its ends. The open circles indicate
the N-W disk points. Thus the  N-W region has a depth similar to the central region of the bar. 
The plot also suggests that the eastern end of the bar is more flared than the western end. The errors
shown are the standard deviation of the average, thus a large error indicates a large range in the 
depth values.\\

We also used the MCPS data to estimate the depth in the bar region. This data set could not be corrected
for data incompleteness and thus might underestimate the RC stars, especially in the crowded bar region.
This is also reflected in the number counts, since a larger area is required to obtain a similar number
of red clump stars. We used larger area bins for the MCPS data and estimated the depth.
The colour-coded figure for the LMC (bar and the disk) is shown in the upper panel of figure 9. 
A prominent feature of the plot is the lop-sided distribution of the red dots (greater depth)
 when compared to the black dots (lesser depth). Thus, this plot reveals that 
there is a variation in depth across the disk of the LMC.
The average depth for the bar estimated based on 320 regions is 4.3 $\pm$ 1.0 kpc, which is very similar
to the value estimated from the OGLE II data. 
The large dispersion is not due to errors in the estimation of the depth of
individual regions, but is due to the presence of regions with varying depths in the bar region.\\

We used the MCPS data to estimate the depth of the disk. We considered regions outside the location
of the bar as disk regions. The estimated depth ranges from  650 pc to 7.01 kpc.The minimum value of depth is limited by errors. The fraction of such regions, where the minimum value is limited by errors, is 0.44\%. 
The average of the disk alone is estimated to be 3.44$\pm$1.16 kpc. 
The average values of the depth for different disk regions were estimated and are tabulated in
table 1, along with the values for the bar for comparison. It can be seen that the
average depth of the northern disk is greater than the southern disk by more than 2 $\sigma$ (of the
southern disk). The depth of the northern disk is
similar to the depth estimated for the bar. Thus the bar and the northern disk of the LMC have
greater line of sight depth, whereas the east, west and the southern disk have reduced depth. 
The variation of depth as a function of RA (bottom panel) and Dec (top panel) is shown
in figure 11. The depth variation as a function of RA indicates that the east and the west ends have
a depth less than the bar, whereas the depth variation as a function of Dec indicates that the
depth reduces from north to the south disk, with increased depth in the bar regions. The fact that
the northern disk of the LMC has a greater depth compared to the other regions seems to be a surprising
result. It will be interesting to study the line of sight depth of regions located further north, to 
find out how far this trend continues in the disk. Figure 9 (upper panel) also 
gives a mild suggestion that
the depth of the disk gradually reduces towards the south, especially on the south-western side.
This is indicated by the increase of yellow points and the appearance of cyan points.
The plot also mildly suggests that the maximum gradient in the depth is seen from the
north-east to the south west of the LMC disk. This is similar to the position angle of the
minor axis of the LMC (major axis $\sim$ 120$^o$, van der Marel et al. \cite{van02}). 
Within the radius of the disk studied here, there is no evidence for flaring of the disk. An estimation
of depth of outer disk at large radii is essential to confirm the above indicated trends.\\

\section{Results : The SMC}
In the case of the SMC also, we used OGLE II and MCPS data sets. Similar to the LMC, the area
covered by OGLE II is mainly the bar region, whereas the bar and disk are covered by the
MCPS data.
The depth of 150 regions (OGLE II data) and 600 regions (MCPS data) of the SMC were calculated.\\

A colour coded, two dimensional plot of depth for these two data sets are shown in figure 13 (OGLE II data in the lower panel
and MCPS data in the upper panel). The optical center of the SMC is 
taken to be RA = 00$^h$ 52$^m$ 12.5$^s$ , Dec = -72$^0$ 49' 43" (J2000, de Vaucoulers and Freeman, 
\cite{dVF73}). There is no indication of a variation of depth across the disk as indicated by the
uniform distribution of the red and black dots. The prominent feature in both the plots is the
presence of blue and green points indicating increased depth, for regions located
near the SMC optical center. The OGLE II data cover only the bar region and it can be
seen that this data is not adequate to identify the extension of the central feature, whereas
the MCPS data clearly delineates this feature.\\

The net dispersions range from 0.10 to 0.35 mag (a depth of 2.8 kpc to 9.6 kpc) in 
the OGLE II data set and from 0.025 mag to 0.34 
mag (a depth of 670 pc to 9.47 kpc) in the MCPS data set. The minimum depth estimated in the MCPS 
data is limited by errors. The fraction of such regions where the minimum value is limited by errors is 2.83\%. The average value of the SMC 
thickness estimated using the OGLE II data set in the central bar region is 4.9$\pm$ 1.2 kpc and 
the average thickness estimated using MCPS data set, which covers a larger area than 
OGLE II data, is 4.42 $\pm$ 1.46 kpc. The average depth obtained for the bar region alone
is 4.97 $\pm$1.28 kpc, which is very similar to the value obtained from OGLE II data. The depth
estimated for the disk alone is 4.23$\pm$1.47 kpc. Thus the disk and the bar of the SMC do
not show any significant difference in the depth. The marginal difference between the bar and
the disk depths is due to the presence of higher depth regions near the center. Thus, except
for the central feature, the depth across the SMC appears uniform.
Our estimate is in good agreement with the depth estimate of the SMC using eclipsing
binary stars by North et al. (\cite{N08}). They estimated a 2-sigma depth of 10.6 kpc,
which corresponds to a 1-sigma depth of 5.3 kpc.\\ 


In order to study the variation of depth of the SMC (OGLE II data) along the RA,
dispersion corresponding to the depth is plotted against RA in figure 14.
The lower panel shows all the regions along with the error in depth estimation for each location.
The upper panel shows the depth averaged along Dec and the error indicates the standard deviation
of the average. Both the panels clearly show the increased depth near the SMC center. There is no
significant variation of depth along the bar.\\

For MCPS data, the dispersion corresponding to depth is plotted against RA as well as DEC in figure
15. There is an indication of increased depth near the center, as seen before. 
The plot also indicates that there is no significant variation in depth between the bar
and the disk, and there is no indication of variation of depth across the disk. In figure 16, the 
depth averaged over RA and Dec are shown in the upper and lower panel respectively. These are  
plotted for a small range of Dec ($-$72.0 - $-$73.8 degrees) and RA (10 - 15 degrees), 
in order to identify the increased depth in the central region.The increased depth near the 
center is clearly indicated. Thus, the depth near the center is about 9.6 kpc, which is twice 
the average depth of the bar region (4.9 kpc). Thus, the SMC has a more or less uniform depth 
of 4.9 $\pm$1.2kpc over bar as well as the disk region, with double the depth near the center.\\

\begin{figure*}
\resizebox{\hsize}{!}{\includegraphics{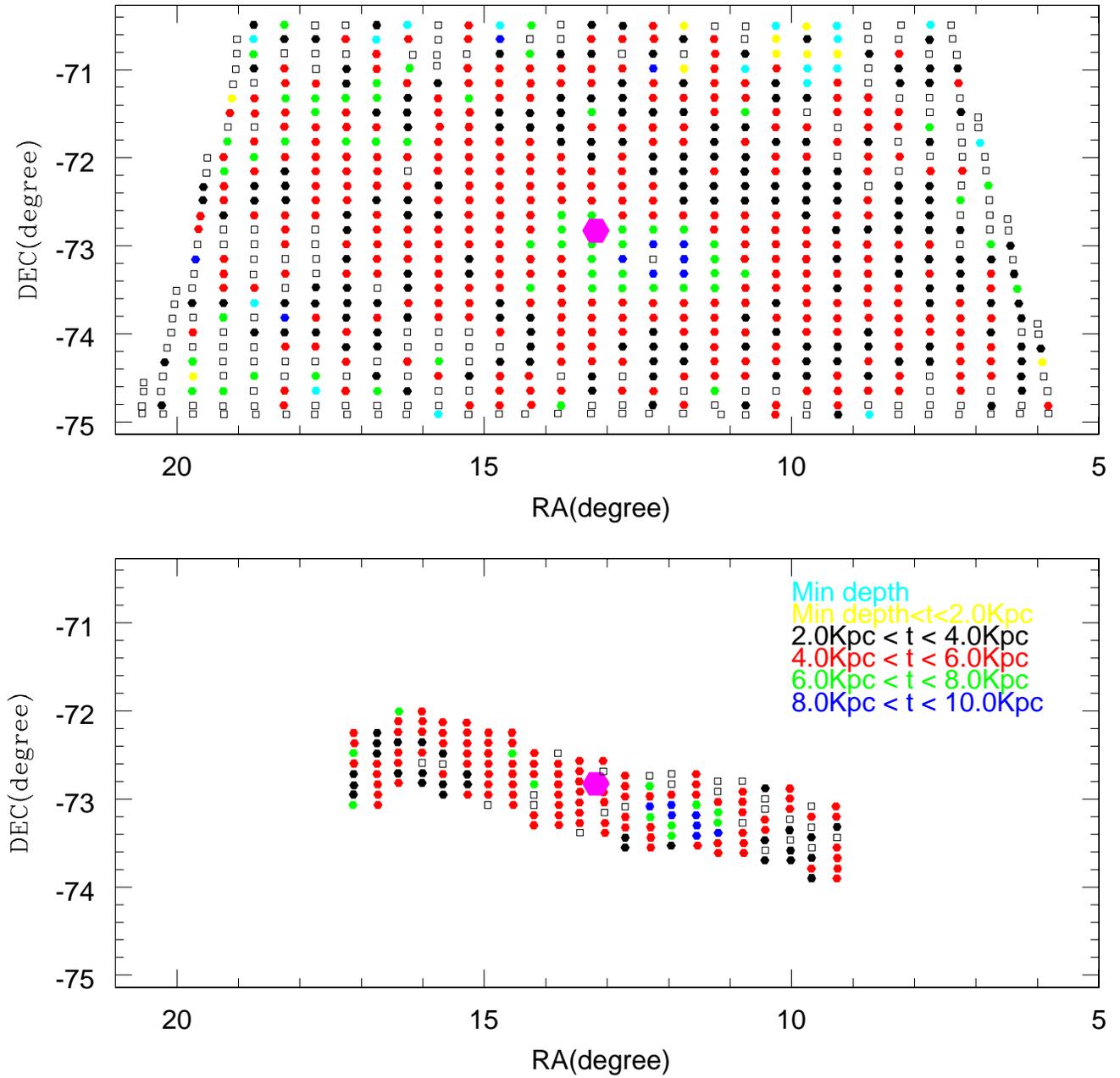}}
\caption{Two dimensional plot of depth (t) in the SMC. Upper panel is for the MCPS data and lower panel 
is for OGLE II data.The colour code is same for both the panels. The magenta dot represents the optical center of the SMC.
The empty squares represent the omitted regions with poor fit.}
\end{figure*}

\begin{figure}
\resizebox{\hsize}{!}{\includegraphics{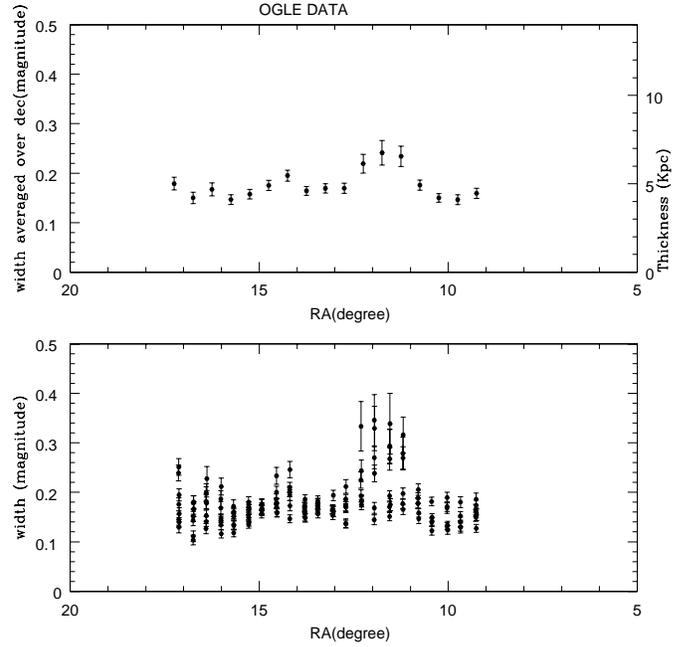}}
\caption{Lower panel: Width corresponding to depth against RA for bar region of
the SMC (OGLE II data). Upper panel: Average of depth along the declination against 
RA in the bar region of the SMC (OGLE II data).}
\end{figure}

\begin{figure}
\resizebox{\hsize}{!}{\includegraphics{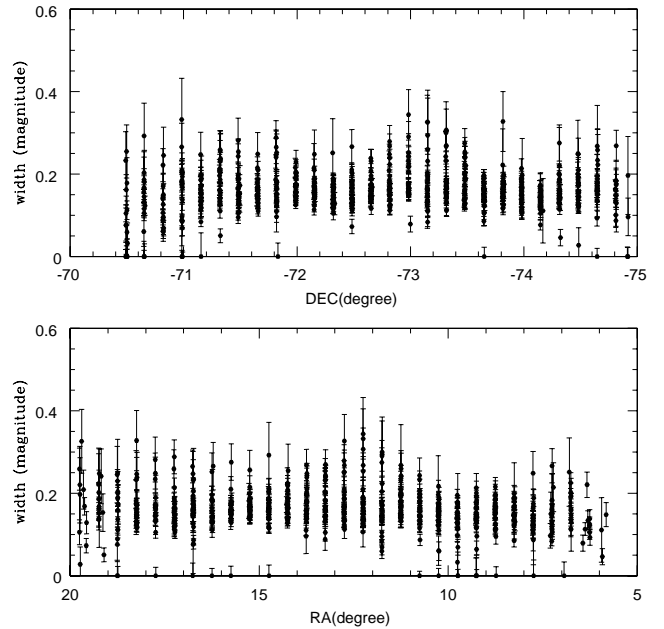}}
\caption{Width corresponding to depth against RA in the lower panel and against Dec in the upper panel 
for the SMC (MCPS data).}
\end{figure}

\begin{figure}
\resizebox{\hsize}{!}{\includegraphics{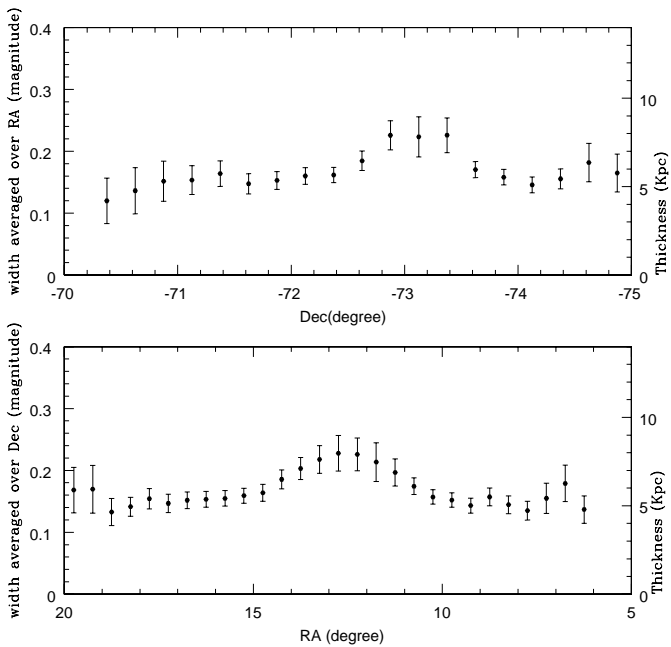}}
\caption{Lower panel: Width corresponding to depth averaged over Dec and plottedagainst RA for a small range of Dec in the central region of the SMC (MCPS data). Upper panel: Width corresponding to depth averaged over RA and plotted against Dec for a small range of 
RA in the central region of the SMC (MCPS data).}
\end{figure}

\begin{figure}
\resizebox{\hsize}{!}{\includegraphics{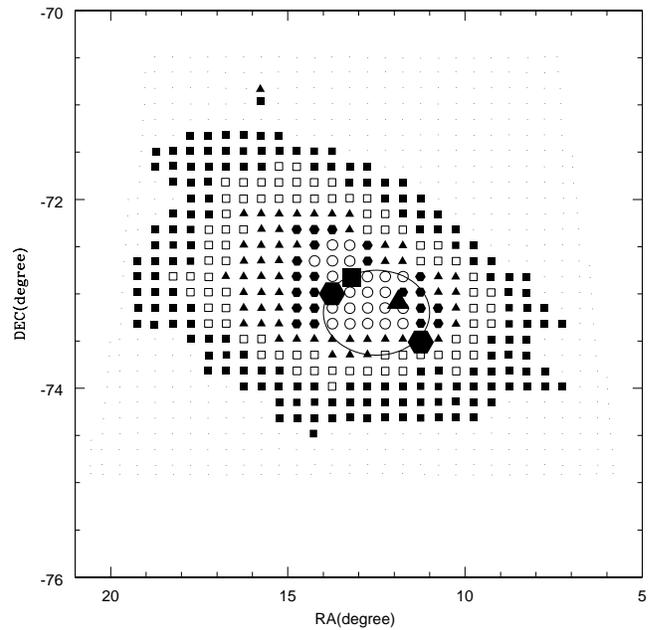}}
\caption{Two dimensional plot of density distribution estimated from MCPS data. The small open circles
in the central region indicate the high density regions. The ellipse shows the boundary of regions
with large depth, the large hexagons indicate the stellar peaks found by Cioni et al. (\cite{C00}), the
large triangle indicate the HI peak (Stanimirovic et al. \cite{SS04}) and the large square denotes
the optical center.}
\end{figure}

\begin{table*}
\centering
\caption{Depths of different regions in the LMC \&\ SMC. These are line of sight depths and 
need to be corrected for inclination, to estimate the actual depth.}
\label{Table:1}
\vspace{0.25cm}

\begin{tabular}{lrrr}
\hline \\
Region & Range of depth (kpc) & Avg.depth (kpc)& Std.deviation (kpc)\\ \\
\hline
\hline \\

LMC eastern bar (RA $>$ 84$^o$.0) & 0.69-9.10 & 4.95 & 1.49 \\

LMC western bar (RA $<$ 80$^o$.0) & 1.26-10.44 & 4.14 & 1.35 \\

LMC central bar (80$^o$.0 $<$ RA $<$ 84$^o$.0)& 0.69-8.50 & 3.21 & 1.03 \\ \\

LMC bar average & 0.69-10.44 &  3.95 & 1.42\\ \\
\hline
\hline  \\

LMC eastern disk (RA $>$ 88$^o$.0, $-$68$^o$.0 $>$ Dec $>$ $-$71$^o$.0) & 0.65-5.89 & 2.80 & 0.92\\

LMC western disk (RA $<$ 74$^o$.0, $-$68$^o$.0 $>$ Dec $>$ $-$71$^o$.0) & 0.65-5.58 & 3.08 & 0.99\\

LMC northern disk (Dec $>$ $-$68$^o$.0) & 1.00-7.01 & 4.17 & 0.97\\

LMC southern disk (Dec $<$ $-$71$^o$.0)& 0.65-4.58 & 2.63 & 0.79\\ \\

LMC disk average & 0.65-7.01 & 3.44 & 1.16\\ \\
\hline
\hline \\

SMC bar & 3.07-9.53 & 4.90 & 1.23 \\

SMC disk & 0.67-9.16 & 4.23 & 1.48 \\\\
\hline
\end{tabular}

\end{table*}

\section{Discussion}
The line of sight depth of MCs are estimated using red clump stars as tracers. They are intermediate age 
stars with ages greater than 1 Gyr, hence, the depth estimates correspond to the intermediate age disk.
The depth of the intermediate age disk may give clues to its evolution, as a thin disk is indicative
of an undisturbed disk, whereas a thick disk would indicate a heated up and hence disturbed disk.
Thus the depth of the disks as well as the bar regions of these galaxies are clues
to the evolution of these galaxies. The analysis presented here 
estimates the dispersion
in the RC magnitude distribution due to depth, after correcting for dispersion due to other effects.
The corrections due to the internal reddening and observational error were estimated for each region
and were corrected accordingly. 
The correction for the presence of an age and metallicity range in the RC population, along with
star formation history is done using the dispersion estimated by Girardi \& Salaris (\cite{GS01}).
Thus the values estimated here have a bearing on the assumptions made during the correction
of the above effect.\\

The variation in the estimated dispersion (after all the corrections) is 
assumed to be due to variations in depth across the galaxy. If this is actually due to
the differences in RC population, as a result of variation in age, metallicity and star 
formation history, then the results indicate that the RC population and their properties are
significantly different in the bar, north and south of the disk, for the LMC. Recent studies show that 
the above parameters are more or less similar across the LMC (Subramaniam \& Anupama (\cite{SA02}), 
Olsen \& Salyk (\cite{OS02}) and van der Marel \& Cioni (\cite{vc01})). 
Variation of star formation history and metallicity across the LMC has been
studied by Cioni et al. (\cite{C06LMC}), Carrera et al. (\cite{Car08}). They found
small variations in the inner regions, but large variations in the outer
regions. Cioni et al. (\cite{C06LMC})
mapped the variation of star formation history as well as metallicity using the AGB stars. They
found that the south and south-western regions could be metal poor whereas
the north and north-eastern regions could be metal rich. Regions near the
Shapley constellation III were found to have a younger population. This region is
close to the northern limit studied here. Piatti et al. (\cite{Piatti99}) and Dolphin (\cite{D00}) found that the RC
population in the far northern regions is structured. All the above regions with varying RC population
are outside the regions studied here. 
The results presented in this study suggests that the northern regions and the
bar have large dispersion, probably due to depth when compared to the east, west and the southern regions.
If this is not due to increased depth in these regions, then it would mean
that the stellar populations in these regions are significantly different. In
either case, the northern disk and the bar seem to be different from the rest
of the LMC disk.\\

As incompleteness correction is done in one data set (OGLE II ) and not in the other (MCPS)
 we compared the depth estimates before and after adopting the completeness correction. We
found that the change is within the bin sizes adopted here.
Thus, incorporating the incompleteness correction has not changed the
results presented here. The incompleteness correction is large in the central regions of the LMC bar,
where the correction is between 30-40 percent. The outer regions of the
bar (6-15\%) as well as the NW region (5-9\%) have less correction. The outer regions
are thus less affected by the incompleteness. Thus the incompleteness problem
is unlikely to affect the MCPS RC distribution in the LMC disk, whereas it may be
unreliable in the central regions of the LMC bar. In the SMC, the incompleteness correction in the central regions is about 12\% and that in the outer region is about 5\%. This is the case for SMC also. The 
incompleteness in the MCPS data does not
affect the results presented here.\\

We have removed regions in the MCs with poor fit as explained in section 4.
These regions are likely to have different RC structures suggesting a large variation
in metallicity and/or population. The fraction of such regions is about 8\% in the LMC and 5.3\% in the SMC.
Such regions are indicated in figures 9 \& 11.
Thus to a certain extent, the above procedure has eliminated the regions with
very different metallicity and star formation history that are seen in
most of the regions. Apart from the above, the remaining regions studied here might have some variation
in the the RC population contributing to the depth estimated.
The results presented in this study will include some contribution from the population
effect.\\
 
The estimated depth for various regions in the LMC is given in table 1. These values 
correspond to the 1-sigma depth. In the case of the
LMC, the line of sight depth estimated here is for the inclined disk. To estimate the
actual depth of the disk, one needs to correct
for the inclination. Assuming the inclination to be 35 degrees (van der Marel et al. \cite{vc01}), 
the actual depth of the bar is 4.0$\times$cos{\it(i)} = 3.3 $\pm$ 1.0 kpc. 
Similarly, the southern disk has an actual depth of about 2.2 $\pm$ 1.0 kpc, whereas
the northern disk is similar to the bar. Note that, after correcting for inclination, the
depths in the northern and the souther regions are within the errors. This is because the difference 
in depth reduces due to correction for inclination, but the error does not.
These values should be used when one compares the depth
of the LMC with that of other galaxies. Thus, the LMC bar and the disk are thicker than the thin disk
of our Galaxy ($\sim$ 100 pc). 
The scale height of the bar could be taken as half of its depth, assuming that the
bar is optically thin. 
The z-extent of the bar, which is the scale height (1.65 kpc), is found to be 
similar to its scale length (1.3 - 1.5 kpc, van der Marel \cite{van01}). 
Hence the bar has a depth similar to its
width. Thus the bar continues to be an unexplained structure/component of the LMC.\\ 

The LMC bar is found to be fairly thick (line of sight depth = 4.0$\pm$1.4 kpc).
We also find evidence for flaring and a disturbed structure at the ends of the bar region. 
The thick and flared bar of the LMC indicates that this region of the LMC is perturbed. 
The structure of the LMC bar as delineated by the RC stars showed warps (Subramaniam \cite{S03}), which is  also
clear indication of disturbance. 
The depth estimates also suggest that the LMC disk is fairly thick (2.4 - 4.0 kpc), with a decrease
in depth/thickness and/or varying stellar population from the north to the south. 
The tidal effects due to LMC-Galaxy interactions (if they were interacting) are unlikely 
to cause this, as
the tidal effects are stronger near the outer regions and weaker towards the inner regions. Flaring
of the disk is expected if tidal interactions are present. On the other hand,
except for the thicker northern region, flaring of the disk is not seen, at least up to the radii
studied here. Hence we do not see any
evidence for tidal interactions, at least in the inner disk.  The recent results on 
the proper motion of the LMC and SMC (Kallivayalil et al \cite{K06a}) suggested that the Clouds are approaching our
Galaxy for the first time. This would suggest that the LMC has not interacted with our Galaxy
before. Our results are in good agreement with this scenario.\\

In general, thicker and heated up disks are considered as signatures of 
minor mergers (Quinn \& Goodman \cite{QG86}; Velazquez \& White \cite{VW99}).
Thus, the LMC is likely to have experienced minor mergers in its history. 
The presence of warps in the bar (Subramaniam \cite{S03})
and evidence of counter rotation in the central regions (Subramaniam \& Prabhu \cite{SP05}) also support
the minor merger scenario. Thus, it is possible that the LMC has experienced minor mergers
during its evolution. These mergers have affected the northern disk and the bar. The variation
in depth observed across the LMC disk could constrain the way in which these mergers could
have happened.\\

The SMC is found to have a depth greater than the LMC. The disk and the bar does not show 
much difference in depth.
A striking result is the increased depth in the central region of the SMC. The profile 
of the depth near the
center (figure 16) looks very similar to a typical luminosity profile of a bulge. This could 
suggest the presence
of a bulge near the optical center of the SMC. If a bulge is present, then a density/luminosity 
enhancement in this region is also expected. We plotted the observed stellar density in each region
from the MCPS data to see whether there is any such central enhancement. This is shown in 
figure 17. The regions with high density are shown as open circles, located close to the optical center. 
The regions with large depth are found to be within the ellipse
shown in the figure. It can be seen that regions with  highest stellar density lie more or less
within this ellipse.
Cioni et al. (\cite{C00}) studied the morphology of the SMC using the DENIS catalogue. They
found that the distribution of AGB and RGB stars show two central
concentrations, near the optical center,
which match with the carbon stars by Hardy et al. (\cite{Hardy89}). They also found
that the western concentration is dominated by old stars. The approximate locations
of these two concentrations found by Cioni et al. (\cite{C00}) are shown as hexagons in figure 17. 
Also, the strongest HI concentration in the SMC map by Stanimirovic et al. (\cite{SS99}) falls between these
two concentrations. The maximum HI column density, 1.43 $\times$ 10$^{22}$ atoms cm$^{-2}$
is located at RA = 00$^h$ 47$^m$ 33$^s$, Dec = -73$^0$ 05' 26" (J2000.0) 
(Stanimirovic et al. \cite{SS04}). This location is shown as a large triangle in figure 17. The
optical center of the SMC is shown as a large square. All these peaks as well as the
optical center are located on or within the boundary of the ellipse.
Thus, the peaks of stellar as well as the HI density are found within the central region with large depth.
This supports the idea that a bulge may be present in the central SMC. 
This bulge is not very luminous, but clearly shows enhanced density. 
It is also the central region of the triangular shaped bar.\\ 

The increased dispersion near the SMC center, which is interpreted as due to large depth, could
be partially due to the presence of RC population which is different. 
Cioni et al. (\cite{C06SMC}) did not find any different
population or metallcity gradient near the central regions.
Tosi et al. (\cite{T08}) obtained deep CMDs of 6 SMC regions to study the
star formation history. Three of their regions are located close to the bar and three are outside the bar.
They found an apparent
homogeinty of the old stellar population populating the subgiant
branch and the clump. This suggested that there is no large differences
in age and metallicity among old stars in these locations.
Their SF1 region is located close to the region of large
depth identified here. The RC population in this region is found to be
very rich and the spread in magnitude is greater than those found in the
other CMDs. This spread is also suggestive of increased depth near
this location. It will be worthwhile to study the star formation history of regions near the
SMC center to understand how different the stellar population is in this
suggested bulge.\\

It may be worthwhile to see whether this bar is actually an extended/deformed bulge.
It is interesting that the so-called triangular 
shaped bar of the SMC is also an unexplained component, which does not show the 
signatures of a typical bar. This could naturally
explain the formation of the odd shaped bar in the SMC. Thus, we propose that the central SMC has
a bulge.
The elongation and the rather non-spherical appearance of the
bulge could
be due to tidal effects or minor mergers (Bekki \& Chiba \cite{BC08}).\\

\section {Disk and halo of the LMC}
Subramaniam (\cite{S06}) studied the distribution of RR Lyrae stars in the bar region of 
the LMC. She found that
the RR Lyrae stars in the bar region have a disk-like distribution, but halo-like location.
The RR Lyrae stars are in the same evolutionary state as the RC stars, except that the RR Lyrae
stars belong to an older and metal poor population. Therefore they are good tracers of the halo.
Thus, it will be interesting to compare the depth of the halo as defined by the RR Lyrae stars
and the depth of the disk as defined by the RC stars.\\

Subramaniam (\cite{S06}) derived the dispersion in the extinction corrected average I magnitude 
of RR Lyrae stars in
the bar region. After correcting  for contribution to the dispersion due to other factors, the
dispersion due to depth was estimated. The total depth estimated for RR Lyrae stars in this paper
was compared with the RC depth. 
The upper panel of figure 18 shows the dispersion
as estimated from RR Lyrae stars as open circles and that estimated average from RC stars as dots.
The figure shows that the RR Lyrae depth ranged between 4.0 and 8.0 kpc (corresponding to a 
scale height of 2.0 -- 4.0 kpc, as reported in the above paper). 
Since the RR Lyrae stars were studied in the bar region, this comparison is valid only
for the bar region. It is assumed that the bar is part of the disk, and hence allowing the comparison of
disk vs halo. It can be seen that the depth as indicated by the RC stars is approximately
the lower limit set by the RR Lyrae stars. That is, the RR Lyrae stars span a greater depth
than the RC stars. Thus, at least in the central region of the LMC, the halo, as delineated by
the RR Lyrae stars, has a much greater
depth than the disk, as delineated by the RC stars. This supports the idea that there is an
inner halo for the LMC. It is interesting that, in an outside-in collapse scenario, the
disk starts to form at the end of the halo formation. This transition is more or less
indicated in the figure as the transition where the RR Lyrae stars stop forming in the halo
and the RC stars take over and the disk forms. In order to make this statement conclusive, 
we need to make such a comparison
for the entire  disk region of the LMC, not just the bar region. The present study, though
indicative, suggests that the halo to disk transition of the LMC follows the outside-in
formation, at least in the inner regions.\\

\begin{figure}
\resizebox{\hsize}{!}{\includegraphics{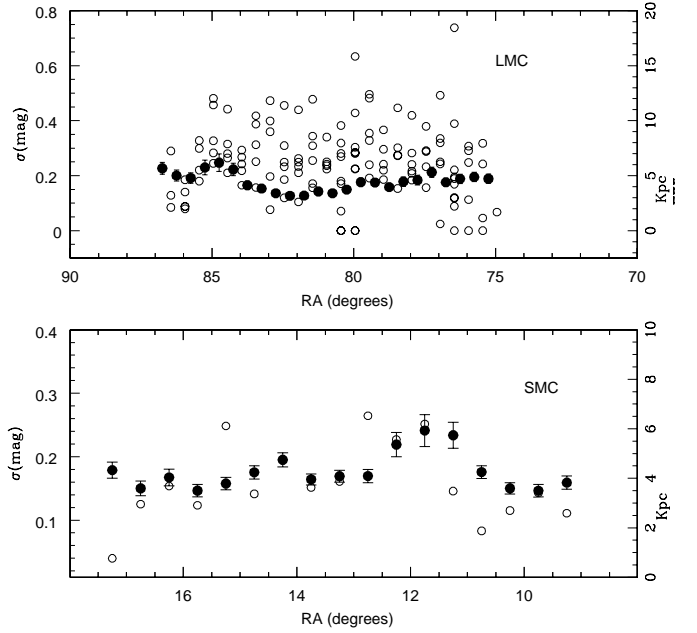}}
\caption{Lower panel: Width corresponding to depth estimated from RR-Lyrae (open circles)
and from red clump stars (closed circles) is plotted against RA for the SMC. Theupper panel shows a similar plot for the LMC.}
\end{figure}

\section {Disk and halo of the SMC}
A similar comparison can be made between the halo and disk/bar of the SMC. The RR Lyrae
stars from the OGLE II data were analysed similarly to the procedure adopted by
Subramaniam (\cite{S06}).
The depth along the line of sight is estimated from the observed
dispersion in the extinction-corrected mean I magnitude of 458 ab type
RR Lyrae stars. The dispersion due to depth alone was estimated after correcting for the
metallicity and evolutionary effects. These were compared
with the dispersion estimated from RC stars in the lower panel of figure 18. 
It can be seen that, contrary to what is
seen in the case of the LMC, both populations show a very similar dispersion in the SMC.
The figure not only suggests that the RR Lyrae stars and the RC stars occupy a similar depth, but
also indicates that they show a similar depth profile across the bar. The increased depth near the
optical center is also closely matched. 
This suggests that the RR Lyrae stars and RC stars are born in the same location
and occupy a similar volume in the galaxy.  This is a puzzling combination, as in general, 
the RR Lyrae stars and RC stars
do not co-exits, as they belong to the halo and disk population respectively.\\

On the other hand, there is one location in the galaxy where this can take place, which is the bulge.
That is, we do see the very metal-poor low mass stars (RR Lyrae stars) and
the higher mass metal rich counter parts (RC stars) co-existing in the bulge. Thus, the co-existence
and the similar depth of RR Lyrae stars and the RC stars in the central region of the SMC
can be easily explained, if it is the bulge. Thus the bar region of the SMC could in fact
be a bulge. This is in good agreement with the result obtained earlier, where a bulge-like
depth, and enhanced stellar and HI density were found near the optical center. 
If this is true, then most of the RR Lyrae stars in the central region belong to this
bulge and not the halo. The depth of both these populations in the outer regions need to
be compared to make this picture complete.\\

Thus, the formation and evolution of the two clouds do not seem to be similar. The LMC is more or less
an irregular galaxy with a disk. On the other hand, the SMC could be a spheroid. In this study
we propose that the SMC has a bulge and the so-called bar is possibly this 
deformed/extended bulge. The LMC does not
have a bulge, but, the SMC has managed to form a bulge. The LMC seems to have
undergone minor mergers, whereas the SMC seems to have experienced tidal forces and/or minor mergers. 
The structure of the inner LMC agrees well with the outside-in formation of the LMC.
 We do not find any evidence for such an inner halo in the SMC.
Thus, even though the clouds are located close to each other now, the early formation and 
evolution of these two galaxies appear different.\\

\section {Conclusions}
1. The LMC and the SMC are found to have large line of sight depths (1-sigma) for the bar 
(4.0$\pm$1.4 and 4.9$\pm$1.2 kpc) and disk (3.44$\pm$ 1.16 kpc and 4.23$\pm$1.48 kpc).\\
2. The LMC bar(4.0$\pm$1.4 kpc) and the northern disk (4.17$\pm$0.97 kpc) have similar, but large depth. The eastern (2.8$\pm$0.92 kpc), western (3.08$\pm$0.99 kpc) and the southern (2.63$\pm$0.79 kpc) disk have similar, but reduced depth. This may also be interpreted as due to different stellar populations.\\
3. The depth profile indicates flaring of the LMC bar.\\
4. The LMC halo is found to have greater depth than the disk/bar, which
supports the presence of the inner halo for the LMC. 
The structure of the inner LMC agrees well with the outside-in formation of the LMC.\\
5. The SMC bar and the disk have similar depth, with no significant depth variation across the disk. \\
6. Increased depth is found near the optical center of the SMC.\\
7. The co-existence of RR Lyrae stars and RC stars in the central volume, along with the
increased depth and stellar \& HI density near the center, suggest that the SMC possibly 
has a bulge. The central bar may be this deformed/extended bulge.\\ 
8. The large depths of the L\&SMC suggest that they have experienced heating, 
probably due to minor mergers.\\

Acknowldegements: We thank the anonymous referee for useful comments which improved the presentation of the paper.

\end{document}